\documentclass[twocolumn,prl,showpacs,preprintnumbers,amsmath,amssymb]{revtex4}

\usepackage{graphicx}
\usepackage{dcolumn}
\usepackage{bm}

\begin{document}


\title{Metallic behavior of lightly-doped La$_{2-x}$Sr$_x$CuO$_4$ with ``arc" of Fermi surface}

\author{T. Yoshida$^{1,2}$, X. J. Zhou$^2$, T. Sasagawa$^3$, W. L. Yang$^4$,
P. V. Bogdanov$^2$, A. Lanzara$^{2,4}$, Z. Hussain$^4$, T.
Mizokawa$^1$, A. Fujimori$^1$, H. Eisaki$^2$, Z.-X. Shen$^2$, T.
Kakeshita$^3$, S. Uchida$^3$} \affiliation{$^1$Department of
Physics and Department of Complexity Science and Engineering,
University of Tokyo, Bunkyo-ku, Tokyo 113-0033, Japan}
\affiliation{$^2$Department of Applied Physics and Stanford
Synchrotron Radiation Laboratory, Stanford University, Stanford,
CA94305} \affiliation{$^3$Department of Advanced Materials
Science, University of Tokyo, Bunkyo-ku, Tokyo 113-8656, Japan}
\affiliation{$^4$Advanced Light Source, Lawrence Berkeley National
Lab, Berkeley, CA 94720}

\date{\today}

\begin{abstract}
Lightly-doped La$_{2-x}$Sr$_x$CuO$_4$ in the so-called
``insulating" spin-glass phase has been studied by angle-resolved
photoemission spectroscopy. We have observed that a
``quasi-particle" (QP) peak crosses the Fermi level in the node
direction of the $d$-wave superconducting gap, forming an ``arc"
of Fermi surface, which explains the metallic behavior at high
temperatures of the lightly-doped materials. The QP spectral
weight of the arc smoothly increases with hole doping, which we
attribute to the $n \sim x$ behavior of the carrier number in the
underdoped and lightly-doped regions.
\end{abstract}

\maketitle

How the electronic structure evolves with hole doping from the
Mott insulator to the metal/superconductor phase is believed to be
a key issue to elucidate the mechanism of superconductivity in the
high-$T_c$ cuprates. However, the issue has remained controversial
until now. In order to clarify the doping-induced changes of the
electronic properties, one has to understand the electronic
properties in the vicinity of the Mott insulating state and hence
the nature of the ``insulating" spin-glass phase (or diagonal
stripe phase \cite{wakimoto}) and the metal-insulator transition
(MIT) at $x\sim0.06$ separating this phase from the
superconducting phase. Recent transport studies on
La$_{2-x}$Sr$_x$CuO$_4$ (LSCO) have indicated metallic ($d\rho/dT>
0$) behaviors at high temperatures in the spin-glass phase ($0.02
< x < 0.06$), even down to the extremely light doping ($x \sim
0.01$) inside the antiferromagnetic ``insulating" (AFI) phase ($x
< 0.02$) \cite{ando}. In this doping region, the absolute value of
the in-plane electrical resistivity is far above the Mott limit
for metallic conductivity in two-dimensional metal, $k_{\rm F}l
\ll 1$, where $k_{\rm F}$ is the Fermi wave number and $l$ is the
carrier mean-free path. These observations raise a question of the
intrinsic nature of the ``insulating" spin-glass phase and hence
the intrinsic nature of the MIT at $x \sim 0.06$.

In  previous angle-resolved photoemission (ARPES) studies,
underdoped LSCO has shown a ``two-component" electronic structure
around ($\pi,0$), which manifests the evolution of in-gap states
 into the Mott insulator with hole doping \cite{inoPRB}. The
chemical potential pinning in the underdoped region also indicates
the existence of the in-gap states \cite{inoCP}. However, since
the pseudogap around ($\pi,0$) increases with decreasing hole
content, it has not been clear how the in-gap states exist and
contribute to the transport properties of the lightly doped
materials mentioned above.

In this Letter, we have performed an ARPES study of lightly-doped
LSCO in order to address those intriguing questions. The present
results have revealed a weak but sharp ``quasi-particle" (QP) peak
crossing the chemical potential, i.e., the Fermi level
($E_\mathrm{F}$), along the zone diagonal
$\mathbf{k}$=(0,0)-($\pi,\pi$) direction and hence an ``arc" of
Fermi surface in this region. This indicates that the spin-glass
phase is indeed a metal with a novel electronic structure.
Furthermore, we have found that, with increased doping, the QP
peak intensity smoothly increases up to optimum doping. We propose
that this behavior is associated with the $n \sim x$ behavior of
the carrier number \cite{takagi,uchida}, which is one of the most
peculiar properties of the high-$T_c$ cuprates.

The ARPES measurements were carried out at BL10.0.1 of the
Advanced Light Source, using incident photons of 55.5 eV. We used
a SCIENTA SES-200 analyzer with the total energy resolution of
$\sim 20$ meV and the momentum resolutions of 0.02$\pi$ in units
of $1/a$, where $a = 3.8$ \textrm{\AA} is the lattice constant.
High-quality single crystals of LSCO were grown by the
traveling-solvent floating-zone method. The $T_c$ of $x$ = 0.07,
0.10, 0.15, 0.18 and 0.22 samples were 14, 29, 41, 37 and 28 K,
respectively, and $x$ = 0.00, 0.03 and 0.05 samples were
non-superconducting. The samples were cleaved \textit{in situ} and
measurements were performed at 20 K as in the previous studies
\cite{zhouPRL,yoshida}. This means that we have studied the
``insulator"-to-superconductor transition as a function of $x$ at
the fixed low temperature. In the present measurements, the
electric vector $\vec{E}$ of the incident photons lies within the
CuO$_2$ plane, rotated 45$^\circ$ from the Cu-O direction and is
parallel to the Fermi surface segment around the diagonal region.
Since the transition-matrix element is enhanced in the second
Brillouin zone (BZ) with this measurement geometry \cite{yoshida},
the data in the present paper are all taken from the second BZ
although we retain the notation of the first BZ for the sake of
convenience.

\begin{figure}
\includegraphics[width=7.5cm]{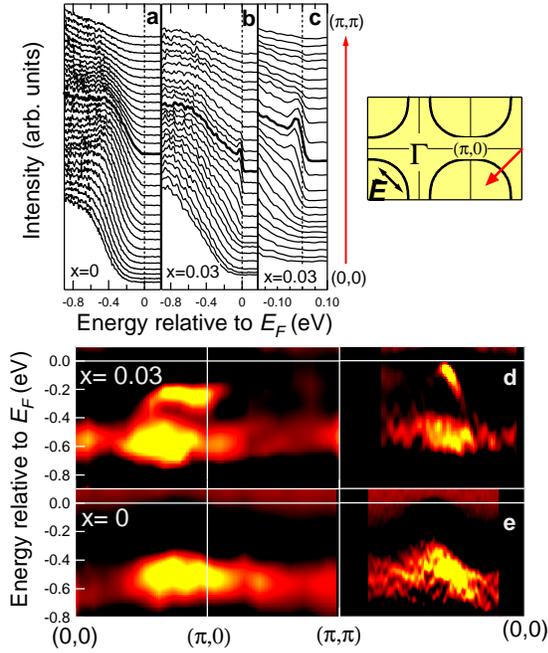}
\caption{\label{nodeEDC}ARPES spectra for LSCO with $x$=0 and
$x$=0.03. Panels a and b are EDC's along the nodal direction
(0,0)-($\pi,\pi$) in the second Brillouin zone (BZ). The spectra
for $x$ = 0.03 are plotted on an enlarged scale in panel c. Panels
d and e represent energy dispersions deduced from the second
derivative of the EDC's. (For detail, see text.)}
\end{figure}

\begin{figure}
\includegraphics[width=9cm]{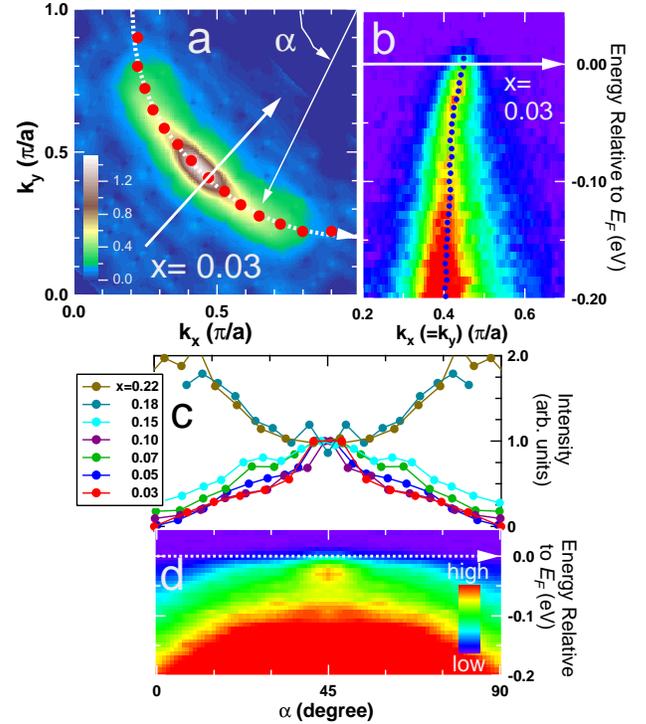}
\caption{\label{nkprofile}a) Spectral weight at $E_\mathrm{F}$ for
$x$=0.03 plotted in the momentum space. b) Spectral intensity in
the energy-momentum ($E$-$\mathbf{k}$) space along the nodal cut
with the peak position of MDC's for $x$=0.03. c) Intensity profile
along the ``arc"/Fermi surface as a function of Fermi angle
$\alpha$ (defined in panel a) for various doping levels. The
spectral intensities have been normalized at $\alpha = 45^\circ$.
d) Spectral intensity in the $E$-$\mathbf{k}$ space along the
``arc" for $x$=0.03. The plots in panels (a), (c) and (d) have
been symmetrized with respect to $\alpha = 45^\circ$.}
\end{figure}

Figure~\ref{nodeEDC} a and b show energy distribution curves
(EDC's) for the $x$ = 0 and 0.03 samples in the (0,0)-($\pi,\pi$)
direction, i.e., in the nodal direction of the $d$-wave
superconductor, where the $d$-symmetry superconducting gap
vanishes. The EDC's for $x$ = 0 show only one broad dispersive
feature at $\sim$ -(0.4-0.6) eV arising from the lower Hubbard
band, consistent with what has been observed in another parent
insulator Ca$_2$CuO$_2$Cl$_2$ \cite{ronning}. In going from $x$=0
to $x$=0.03, the lower Hubbard band becomes a little broader and
an additional sharp feature crossing $E_\mathrm{F}$ appears,
indicating a metallic behavior (see the expanded plot in panel c).
This ``in-gap" state is reminiscent of the coherent part of the
spectral function as predicted by dynamical mean field theory
(DMFT) calculations \cite{zhang}.

In order to highlight the energy dispersion of that sharp feature,
we show a gray-scale plot of the second derivatives of EDC's in
Fig.~\ref{nodeEDC}d and e. The lower Hubbard band observed for $x$
= 0 at $\sim$ -(0.4-0.6) eV remains almost at the same binding
energy with hole doping. Instead, there appear a sharp peak
feature crossing $E_\mathrm{F}$ in the nodal direction, and a
broad feature around ($\pi,0$) at $\sim$ -0.2 eV corresponding to
the ``flat band" as in the previous report \cite{inoPRB}. Because
of the $\sim$ 0.2 eV gap, the electronic states around ($\pi,0$)
would not contribute to the metallic transport in the normal
state, and only the states around the nodal direction would
contribute to it. It is rather striking to observe such a sharp
peak crossing $E_\mathrm{F}$ for hole doping as small as 3\%,
which is near the boundary between the ``insulating" AFI phase ($0
< x < 0.02$) and the ``insulating" spin-glass phase
\cite{niedermayer} ($0.02 < x < 0.06$).

In Fig.~\ref{nkprofile}a, we have plotted the distribution of
spectral weight at $E_\mathrm{F}$ in the $\mathbf{k}$-space
(therefore the energy integration window $E_\mathrm{F}\pm$10 meV
begin set by the energy resolution) for $x$= 0.03. This plot is
obtained from the spectra in the second BZ and symmetrized with
respect to the (0,0)-($\pi,\pi$) line. The color image was
produced by interpolating the spectral intensity of the several
momentum lines in two-dimensional momentum space. Owing to the
(pseudo)gap around ($\pi,0$), only the nodal region remains strong
in the $E_\mathrm{F}$ intensity map. Figure ~\ref{nkprofile}d
indeed indicates that $E_\mathrm{F}$ crossing occurs only near the
nodal direction, forming an ``arc" of Fermi surface seen in
Fig.~\ref{nkprofile}a. The red dots in Fig.~\ref{nkprofile}a are
peak position of the momentum distribution curve (MDC) at
$E_\mathrm{F}$ which represent minimum gap locus in the
$\mathbf{k}$-space. The white broken line is determined by fitting
a tight-binding Fermi surface to the red dots.

In Fig. ~\ref{nkprofile}c, the intensity profile along the ``arc"
is plotted as a function of Fermi angle $\alpha$ (normalized to
the nodal direction $\alpha=45^\circ$). Each point for $x$=0.03
corresponds to a red dot in Fig.~\ref{nkprofile}a and the same
procedure using MDC is applied to obtain the plots for other
compositions. This figure shows how the intensity profile along
the ``arc" changes with doping. These profiles indicate that,
although for $x > 0.1$ some spectral weight appears in the
($\pi,0$) region ($\alpha \sim 0^\circ$ and $90^\circ$), the
length of the ``arc" does not increase significantly with doping
up to $x = 0.15$. In the overdoped samples ($x > 0.15$), the flat
band at $\sim (\pi,0)$ crosses $E_\mathrm{F}$ \cite{inoPRB} and
therefore the intensity in this region increases. A similar ``arc"
feature has been observed in a previous work in the normal state
of underdoped Bi$_2$Sr$_2$CaCu$_2$O$_8$ (Bi2212) \cite{marshall},
and postulated from phenomenological \cite{norman} and theoretical
$t$-$J$ model perspective \cite{wen}. We emphasize the difference
between our observation and the previous work:  Here, the
intensity of the ``arc" changes as a function of doping at a fixed
low temperature whereas in the previous work \cite{norman} the
length of the ``arc" changes as a function of temperature
associated with the opening of a normal-state gap above $T_c$.

\begin{figure}
\includegraphics[width=7.0cm]{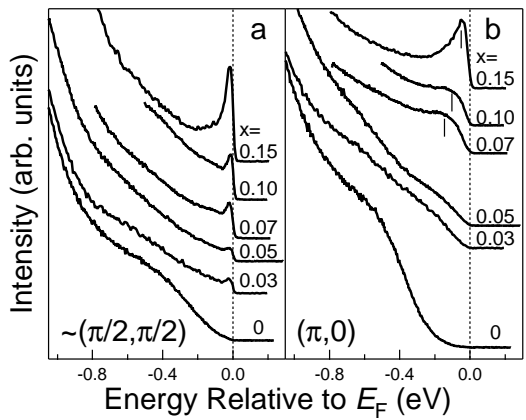}
\caption{\label{EDC}ARPES spectra at $\mathbf{k} =
\mathbf{k}_\mathrm{F}$ in the nodal direction in the second BZ and
those at ($\pi,0$) for various doping levels.}
\end{figure}

Figure~\ref{EDC} shows the evolution of spectra at $\mathbf{k}
\sim (\pi/2,\pi/2)$ and ($\pi,0$) with hole doping. The spectra
have been normalized to the intensity above $E_\mathrm{F}$ which
arises from the high order light of the monochromator. At $\sim
(\pi/2,\pi/2)$, a finite spectral weight exists at $E_\mathrm{F}$
except for $x$ = 0, and increases with $x$ without an abrupt
change across the ``insulator"-superconductor transition boundary
at $x \sim 0.06$. In contrast, the ($\pi,0$) spectra show
(pseudo)gapped behaviors in the underdoped and lightly-doped
regions. We stress that the ``arc" is formed by the ``in-gap"
states of the two component electronic structure as shown in
Fig.~\ref{nodeEDC}.

It has been demonstrated from comparison between LSCO and Nd-LSCO
that the nodal QP is weakened by stripes whereas the flat-band
region remains relatively unchanged \cite{zhouScience,zhouPRL}.
While it remains true that the nodal weight in LSCO is
significantly suppressed when compared with that of Bi2212 taken
under the same condition, it was later found that the nodal
spectral weight can be enhanced in the second BZ
\cite{zhouPRL,yoshida}, indicating that matrix element also played
a role.  Focusing only on the nodal behavior, we have used the
spectra in the second BZ for this analysis. We emphasize here
that, in the underdoped region, states closest to $E_\mathrm{F}$
occur along the nodal direction, as seen from Fig. 3, whereas the
($\pi$,0) region is (pseudo)gapped. In the previous results for
$x$=0.10-0.15, since the gap size was small (less than 10meV), and
the energy integration window was relatively large (30meV), the
spectral weight map included both the nodal states and straight
segment from the flat band around ($\pi$,0) which is considered to
be related to stripes \cite{zhouPRL}. On the other hand, for the
lightly-doped samples presented here, the Fermi arc is clearly
observed because the ($\pi$,0) region is strongly gapped and the
spectral weight plots have been obtained exactly at
$E_\mathrm{F}$.

\begin{figure}
\includegraphics[width=8.5cm]{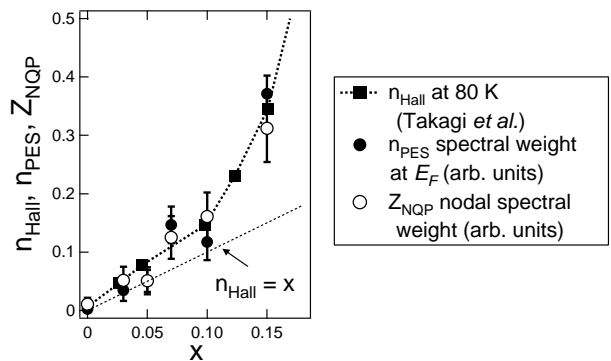}
\caption{\label{Hall}Doping dependence of the nodal QP spectral
weight, $Z_{\rm NQP}$, and the spectral weight integrated at
$E_{\mathrm F}$ over the entire second Brillouin zone, $n_{\rm
PES}$. They show similar doping dependence to the hole
concentration evaluated from Hall coefficient ($n_{\rm Hall}$)
\cite{takagi}.}
\end{figure}

In order to see more quantitatively how the spectral weight at
$E_\mathrm{F}$ evolves with hole doping, we have plotted in
Fig.~\ref{Hall} as a function of $x$ the nodal QP weight $Z_{\rm
NQP}$ defined by the peak intensity of EDC in Fig.~\ref{EDC}a and
the spectral weight at $E_{\rm F}$ integrated over the second BZ
(effectively over the arc region), $n_{\rm PES}\equiv \int
A(\mathbf{k},0)d\mathbf{k}$. Here, $A(\mathbf{k},\omega)$ is the
ARPES intensity normalized as described above. Both $Z_{\rm NQP}$
and $n_{\rm PES}$ monotonically increase with $x$ in a nearly
identical fashion. Such a behavior is consistent with a recent
optical study of lightly-doped LSCO, which has shown that the
Drude weight is finite already in $x$ = 0.03 and smoothly
increases with $x$ in the underdoped region \cite{dumm}. $Z_{\rm
NQP}$ and $n_{\rm PES}$, in the underdoped regime, if properly
scaled, increase in a remarkably similar way to the hole
concentration $n_{\rm Hall}$ derived from Hall coefficients
through $R_H=1/n_{\rm Hall}e$. We therefore tentatively attribute
the $n \sim x$ behavior to the evolution of ARPES spectral weight
at $E_{\rm F}$: $n \sim n_{\rm PES} \sim Z_{\rm NQP}$.

Now we discuss the normal-state transport based on the present
ARPES data for $x$=0.03. From the width $\Delta k$ of the momentum
distribution curve (MDC) at $E_{\rm F}$, the mean free path is
obtained as $l_\mathrm{PES} \sim 1/\Delta k \sim 16$ \AA. One can
also obtain the Fermi velocity $v_{\rm F} \sim 1.5$ eV{\rm \AA}
from the energy dispersion. This yields the electron effective
mass $m^* = \hbar k_{\rm F}/v_{\rm F}$, where the Fermi wave
number $k_{\rm F} \sim 0.62$ \AA$^{-1}$ is measured from
$(\pi,\pi)$, and the mean scattering time of $\tau =
l_\mathrm{PES}/v_{\rm F} \sim 1/v_{\rm F}\Delta k \sim 7$ fs. This
$\tau$ value should be taken as the lower bound for the intrinsic
value because any extrinsic contributions of unknown origin to the
width $\Delta k$ lead to an underestimate of $\tau$. Combining
this with resistivity $\rho = m^*/ne^2\tau= 4.4$ m$\Omega$cm at
$\sim 20$ K \cite{ando}, the upper bound for $n = m^*/e^2\rho\tau$
is given by $\sim 0.04$ per Cu. This is consistent with $n \sim x
= 0.03$ and hence with the conjecture that $n \sim n_{\rm PES}
\sim Z_{\rm NQP} \sim x$. On the other hand, $n \sim 1-x$ leads to
one to two orders of magnitude overestimate. If we apply a
conventional Drude formula of 2D metal to the experimental value
of $\rho$ in the lightly doped region, the large $\rho$ value
yields $k_{\rm F} l \ll 1$, which indicates the apparent breakdown
of the conventional metallic transport \cite{ando}. Note that $l$
in 2D Drude formula is an average value over the Fermi surface
while $l_\mathrm{PES}$ is a local quantity in the momentum space
around node. The present study have shown that the unconventional
$k_{\rm F} l \ll 1$ behavior can be reconciled with the metallic
transport when only a fraction of carriers $n \sim n_{\rm PES}
\sim Z_{\rm NQP}$ contribute to the transport.

Here, it should be mentioned that the resistivity in fact shows an
upturn below $\sim 100$ K, indicating carrier localization. One
possibility is that the localization is caused by the opening of a
small gap in the nodal direction. If we assume that
$\rho\propto\exp(\Delta/k_BT)$with a thermal excitation gap
$\Delta$, one can crudely estimate $\Delta\sim 1-2$meV from the
temperature dependence of $\rho$ in $x$=0.03 sample, which is too
small to observe with the present energy resolution. In fact, the
divergence of $\rho$ at low temperatures does not obey thermal
activation type but is better described by weak localization due
to disorder. The pseudogap behavior (of $\sim$100 cm$^{-1}$) in
the optical spectra \cite{dumm} would be considered as a signature
of such localization.

Finally, we comment on the implication of the present results for
charge inhomogeneity picture. We have shown that the nodal
spectral weight is proportional to $x$. This observation is
consistent with a recent theoretical work of the resonant valence
bond (RVB) picture \cite{paramekanti} as well as an inhomogeneous
picture like charge stripes. As for the nodal state and underlying
Fermi surface around ($\pi,0$), a recent theoretical work have
shown that a metallic nodal state may appear even in an almost
static stripe phase with a slightly irregular inter-stripe
distances \cite{granath}. Those effects may also account for why
the nodal spectra do not show abrupt change from diagonal ($x<$
0.06) to (dynamic) vertical stripe ($x>$ 0.06). Spectroscopic
feature of 1D in the deeply underdoped regime are probably too
subtle to be reflected in our experiment (using twinned samples)
since only weak anisotropy is seen in transport \cite{ando2} and
optical \cite{dumm} experiments. A better understanding of
fluctuations and disorder of stripes probed at different energy
scale are required for reconciling the charge inhomogeneity and a
band-like picture.

In conclusion, we have revealed the origin of the metallic
behavior of lightly-doped LSCO through the observation of a sharp
dispersive QP peak crossing $E_\mathrm{F}$ in the nodal direction,
leading to the formation of an``arc" of Fermi surface. The
spectral weight of the ``arc"  increases with hole doping, which
we propose to be related to the carrier number $n \sim x$ in the
lightly-doped and underdoped regions and the origin of the
unconventional transport with $k_\mathrm{F} l \ll 1$. The present
results combined with the transport data \cite{ando} reveal the
novel metallic state which provides a new scenario of how the the
Mott insulator evolves into a $d$-wave superconductor upon hole
doping.

We are grateful to Y. Ando, D.~N. Basov and N. Nagaosa for
valuable discussions. T. Y. acknowledges the Japan Society for
Promotion Science for their support. This work was supported by a
Grant-in-Aid for Scientific Research ``Novel Quantum Phenomena in
Transition Metal Oxides" and a Special Coordination Fund for the
Promotion of Science and Technology from the ministry of
Education, Science, Culture, Sports and Technology, the New Energy
and Industrial Technology Development Organization (NEDO) and a
US-Japan Joint Research Project from the Japan Society for the
Promotion of Science. ALS is operated by the DOE's Office of BES,
Division of Materials Science, with contract
DE-FG03-01ER45929-A001.

\end{document}